\documentclass[aps,prd,twocolumn,floatfix,nofootinbib]{revtex4}   

\usepackage{amsmath}    
\usepackage{graphicx}   

\newcommand{\ket}[1]{|#1\rangle}
\newcommand{\braket}[2]{\langle #1|#2\rangle}

\begin{document}

\title{How Einstein's quantum hypothesis requires a departure from classical mechanics}
\author{Gabriele Carcassi}
 \affiliation{Brookhaven National Laboratory, Upton, NY 11973}
 \email{carcassi@bnl.gov}
\date{February 15, 2009}

\begin{abstract}
The aim of this work is to show how Einstein's quantum hypothesis leads immediately and necessarily to a departure from classical mechanics. First we note that the classical description and predictions are in terms of idealized measurements that are exact, instantaneous, non-perturbative, independent of each other and process agnostic. If we assume we cannot arbitrarily reduce the strength of a signal, measurements are ultimately perturbative to some degree. We show how a physical description in which the best measurement conceivable, i.e. the ideal measurement, perturbs the system leads to all the concepts present in quantum mechanics including conjugate variables, probabilistic predictions and measurements connected to symmetries.
\end{abstract}

\maketitle

\section{Introduction}


It is unfortunate that, after more than half a century that quantum mechanics has been a core part of our scientific understanding, it is still surrounded by a cloud of mystery and perceived as strange and nonintuitive. It is true that it does predict behavior that is odd and counter to our intuition, but does that have to imply we are bound to feel like something escapes us?

Quantum mechanics is not the only 20th century theory that has strange consequences: the concept of spacetime, time dilation, length contraction, equivalence of mass and energy, curved space and black holes are some of the landmarks of special and general relativity, yet both are usually presented as natural, in fact necessary. We believe that the main difference is that they are presented as coming from a simple physical idea, the invariance of the speed of light in the first case and the equivalence principle in the second, which help us make sense of all the other physical consequences.

Quantum mechanics, with its uncertainty principle, interference and probabilistic predictions, is usually presented as a set of mathematical postulates\cite{sudbery}, usually prefaced by a historical perspective\cite{liboff} or by a heuristic account that gives some sort of justification for them\cite{feynman,griffiths,shankar,sakurai}. We are not told why a Hilbert space must be used as the phase space, or why observables are associated with operators: that is the starting point. The mathematical results derived from the postulates need to be subsequently interpreted physically, with nothing else to connect them together but the mathematical framework. Should the physics not come first? Should the math not be derived from the physics?

We are left to wonder whether we are missing something: what is the "big physical idea" that requires us to abandon the classical description? Maybe if we were to present quantum mechanics derived from it, it would increase our sense of understanding: what is understanding if not being able to identify, in the midst of all that is confusing and misleading, that simple truth from which all others descend?

We are convinced that this idea is something that is already present in all quantum textbooks: Einstein's quantum hypothesis. This states that light consists of and propagates in discrete packets of energy. The aim of this paper is to convince the reader that this assumption, by itself, is sufficient to require departure from the classical description. The language and the level of math used are appropriate for an introductory class, where typically more importance is given to thought experiments and concepts. In fact, the arguments are designed so that they could be used "as is" during the first lecture of an introductory class in quantum mechanics.

In section II we will show how the classical description is in terms of idealized measurements that are exact, instantaneous, non-perturbative, independent of each other and process agnostic. In section III we will show such an idealized measurement is in principle possible using the classical electromagnetic field. In section IV we will show how the introduction of the quantum hypothesis requires all measurements to be perturbative and that this leads to many of the features present in quantum mechanics. In section V we show that the concepts we developed fit extremely well in the mathematical framework. In section VI we extend the arguments to show how the introduction of the equivalence principle and gravitation necessarily leads to a departure from quantum mechanics, as measurements are no longer independent of each other and cannot be regarded as exact or instantaneous. In section VII we go through some common reactions to these arguments, comparing to other types of works.

\section{Ideal measurements in classical mechanics}

Predictions and measurements are fundamental in physics: devising experiments, performing them and comparing their results to our predictions are in essence the activities of a physicist. In this section we are going to review some aspects of these basic concepts in the context of classical mechanics.

In classical mechanics we describe a system by a set of quantities that vary in time. For example, we write $x=x(t)$ for position or $p=p(t)$ for momentum. At each moment in time, we have a prediction for each quantity: if we were to measure, and our description were correct, we would obtain that value. There are a few details, though, that we have to keep in mind.

First we note that while an actual experiment will only measure a quantity within a certain accuracy, the prediction is at least in principle exact. If we increase the accuracy of our measurement, the result will have to be closer to our prediction for the prediction to be correct. The prediction is really for an idealized measurement: one for which the uncertainty is so small that it can be neglected. \footnote{This notion of idealized measurements is not new. For example, J. V. Jos\'{e}, E. J. Saletan, Classical dynamics: a contemporary approach (Cambridge University Press, 1998), p. 6: the principles of classical mechanics "must be understood as statements about idealized experiments". We expand on this notion to identify all the underlying assumptions.}

Secondly, an actual experiment will measure a quantity within a finite interval of time while the prediction is given for an instant. We can imagine that we improve our measurement so that the interval is smaller and smaller. The prediction is again in terms of an idealized measurement, one in which the time interval is so small that it can be neglected.

The third thing that we notice is that the classical description does not require us to say what quantity we are measuring and when: the evolution does not depend on it. In general, an actual experiment will modify the evolution of the system. We can, once again, imagine that we improve our technique so that it affects less and less the future evolution. The prediction is really in terms of an idealized measurement, one in which the perturbation caused by the experiment is so small that it can be neglected.

The fourth point: if we had two or more instances of the same experiment conducted at the same time, or within an interval so small that it could be neglected, the prediction would be the same. While in practice this may be difficult to achieve, ideal measurements do not interfere with each other, so our description does not depend on how many idealized measurements are performed at a particular time: we can have several observers or just one and we obtain the same result.

Fifth and last point: the prediction does not depend on which particular physical process or measurement technique we use for our measurement.

So, when we write $x(t)$ there are quite a few assumptions that go with it. It assumes that, at least in principle, we can imagine an ideal measurement that is exact, instantaneous, non-destructive, independent of other measurements (or "inter-independent") and process agnostic. \emph{This is the best measurement we can conceive.} Our description is given by the outcomes of such idealized measurements at every moment in time. Even when we do describe a "real" measurement, we do so by describing the entire process in terms of these idealized quantities. In other words: we describe the world using the best measurement we can conceive, and from this determine predictions for our real, less perfect, measurement.

It should be noted that if any of those conditions cannot be satisfied conceptually, our description of position as $x=x(t)$ would not make much sense. Also note that since the ideal measurement is non-perturbative, inter-independent and process agnostic, our description does not depend on in what way, how many times, what or even if we measure. \emph{This is what allows us to imagine the outcomes as properties of the system we are studying}: they are going to be the same no matter what we do. But we always have to keep in the back of our mind that even these quantities are the outcome of a process, however idealized it may be.

\section{An ideal process for an ideal measurement}

We have seen how the classical description is in terms of ideal measurements. In this section we identify a physical process that we could conceptually use in classical mechanics to perform such an ideal measurement. Given that we assume that the results do not depend on what process we use, we can choose the process we prefer. For reasons that will become obvious in the next section, we will focus on an ideal position measurement through the use of the electromagnetic field.

In the simplest case, we can think of sending an electromagnetic signal toward our target. The electromagnetic signal will interact with the target, leaving it affected in general. The signal itself will also be changed and when it is received by the detector it will give us some information about the target. In measuring position, we can imagine the signal bounces off the target, changing its momentum a bit, and by knowing the initial and final position and angles of the electromagnetic signal, we can infer the position of our target at the time of impact. Can this process satisfy, under ideal conditions, the requirements of an ideal measurements?

To measure the position with greater accuracy, we can reduce the width of the electromagnetic signal and, ideally, we would make the packet so small that its length can be disregarded. We can also make the duration of the signal as short as we desire, and ideally it will be so short that it can be disregarded. We can make the measurement non-perturbative by lowering the intensity of the signal enough so that the effect on our target can be neglected. And regarding inter-independency, we note that the equations describing the electromagnetic field are linear: different electromagnetic signals of different ideal experiments are not going to interact with each other.

This idealized process can conceptually be used to perform our idealized measurement. This does not in any way prove that the underlying assumptions are right, or that we can actually make such a measurement in practice: it just shows that classical mechanics is consistent. Also note that this example is specifically chosen to remind the reader of thought experiments used for special relativity such as the ideal clock.

\section{The quantum hypothesis and the ideal measurement}

In this section we show how the introduction of the quantum hypothesis no longer allows us to work under the previous idealized conditions. Even an idealized measurement must be perturbative, and we show that this perturbation has to be such that it randomizes another physical quantity, making the future evolution independent of it. The classical description, then, needs to be abandoned for one that takes into account such perturbation and that can only be in terms of probabilistic predictions.

In the previous section we discussed how we can make the ideal measurement non-perturbative by decreasing the intensity of the field. If we introduce the quantum hypothesis, that the electromagnetic field consists of and propagates in discrete packets of energy, we are no longer able to do that arbitrarily: at some point we will reach those units of energy and we cannot go lower than that; a photon is a photon and we cannot divide it. If the target is massive enough so that the effect of one photon can be disregarded, then we can still assume that our ideal measurement is non-perturbative. If that is not the case, then our assumptions regarding our ideal measurement have to change. \emph{The best measurement we can hope to achieve will still affect the system.}

Given that the classical description and predictions are in terms of the non-pertubartive assumption, this means that the quantum hypothesis has more profound effects than one would at first expect. This assumption was the one that allowed us to give a description of the evolution without having to specify what we were measuring and, most importantly, whether we were measuring anything at all: if ideal measurements do not affect the system, we can ignore them. But now even ideal measurements will affect the system, so we will at least need to specify what is it that we are measuring at what moment in time to properly describe the evolution of the system. Now the question becomes whether we can hold onto the other ideal assumptions, and what do we need to sacrifice to keep them.

Ideally we can imagine producing a photon that is extremely well localized both in space and in time, so we can still assume that our ideal measurement is instantaneous and exact. The measurement, as we said, is now perturbative: \emph{measuring one quantity will change at least one other quantity}. So measuring position will need to change some other quantity. Given that ideal measurements are process independent, if measuring position using the electromagnetic field will change momentum, it will have to do so for any other process. Therefore the link between these two quantities is something deep: whenever we describe one we need to describe the other and the effects caused by our idealized measurement.

We still want to assume that our ideal measurement is inter-independent: while the evolution may depend on when and what we measure, it shall not depend on how many times. Let us imagine a time interval $dt$ where we will perform $N$ idealized measurements. Let $x(t)$ and $p(t)$ be the initial position and momentum, and $x(t+dt)$ and $p(t+dt)$ be position and momentum after all measurements. We will need $x(t+dt)$ and $p(t+dt)$ to give us the same description regardless of the value of $N$.

It is clear what needs to happen for the position, the quantity we are measuring: all ideal measurements will have to give us the same precise value, a real number, so $x(t+dt)$ will be simply a real number. It is trickier to consider what needs to happen to the quantity that changes: \emph{each measurement will need to change the momentum, yet the future evolution of the system must not depend on the number of changes}. We can write:
\begin{equation}
\label{measurementEffect}
p(t+dt)=p(t) + \displaystyle\sum_N \Delta p
\end{equation}
where $\Delta p$ is a stochastic variable that describes the change. We assume all changes obey the same distribution, are independent of each other and of the measured position. What we want to understand is the $\Delta p$ distribution: does it give the same value every time? Is it a Gaussian distribution? What is the expectation? What is the variance? Note that for the expectation of $p(t+dt)$ to be independent of $N$ we need the expectation of one perturbation to be the same as many perturbations: this can only happen if the expectation for $\Delta p$ is zero.
\begin{equation}
<\Delta p>=N<\Delta p>
\end{equation}
\begin{equation}
<\Delta p>=0
\end{equation}
Since the change cannot be identical to zero (or the ideal measurement would not be perturbative), the variance needs to be non-zero. The change is then a random change, and does not privilege any direction. This means that both $\Delta p$ and $p(t+dt)$ will necessarily be given in terms of probability distributions. As $N$ increases, $p(t+dt)$ will spread out, given that
\begin{equation}
\sigma^2_{p(t+dt)}=\sigma^2_{p(t)}+N \sigma^2_{\Delta p}.
\end{equation}
The contribution given by $p(t)$ will get smaller, and in the limit of large $N$ $p(t+dt)$ will not depend on $p(t)$ at all. But if this is true for large $N$ it has to be true for any value of $N$, since $p(t+dt)$ can not depend on $N$: \emph{the value of momentum after an idealized measurement of position does not depend on the value of momentum previous to such measurement}! Let us now imagine we start from a slightly different initial condition:
\begin{equation}
p'(t)=p(t) + \delta p.
\end{equation}
Given that $p(t+dt)$ is independent of $p(t)$ is, $p'(t+dt)$ will not change:
\begin{equation}
p'(t+dt) = p(t+dt).
\end{equation}
The final distribution is symmetric under boosts: \emph{after a measurement we will have a symmetry in the conjugate variable, the quantity that changes.} Each value of momentum is therefore equally likely: $p(t+dt)$, as well as $\Delta p$, will be uniform distributions from minus infinity to infinity.

\emph{Our ideal position measurement, then, is going to change momentum so that each value is equally likely.} The way that we can describe processes if we assume that ideal measurements are interactions is quite different: some of our predictions will only be in statistical terms. But we have to realize that this is not just a problem in our description or limited to our measurements.

Not only can we not measure position and momentum precisely at the same time: no electromagnetic process can depend deterministically on both position and momentum or can prepare a state in which both are well defined. If we had such a process, we could use it to make a more precise measurement. But the ideal process we described before represents the best measurement we could make using the electromagnetic force, so such a process cannot exist. As future evolution cannot depend on the value of the momentum at the time we measured position, no process and no measurement will be able to distinguish between those configurations. As what can be distinguished is radically different, we will need to redefine our concept of physical state. This is still the output of a set of idealized measurements, as in classical mechanics, but the output of an idealized measurement is radically different: for each measurement quantity we have a symmetry.

The uncertainty that comes out of assuming ideal measurements are perturbative is not a simple matter of \emph{our} inability to measure precisely (as it is in classical mechanics): it is an intrinsic part of the physical description because electromagnetic processes will depend on that uncertainty, and in fact they do. The stability of the atomic orbits, the tunneling effect, and all the other quantum effects are based, one way or another, on this. And if we assume that ideal measurements and our description do not depend on what process we use, this reasoning extends not only to electromagnetic processes, but to all processes. The scale of this perturbation has to be the same for all forces: it is universal and so is the constant $\hbar$ that defines it.

We have reached the main point of this article: assuming the quantum hypothesis leads necessarily to a physical description that is radically different form the one of classical mechanics. One could go further: we should note that Rovelli\cite{rovelli} derives the whole mathematical framework of quantum mechanics starting from the following postulates
\begin{enumerate}
  \item There exists a maximum amount of relevant information that can be extracted from a system.
  \item It is always possible to obtain new information about the system.
\end{enumerate}
If we define the state as the output of a set of ideal measurements, we imply the first. Noting that one measurement creates a symmetry in another quantity implies the second. Given that these two postulates lead to quantum mechanics, and the quantum hypothesis leads to these two postulates, one could show that the quantum hypothesis, by itself, not only necessarily requires a departure from classical mechanics, but leads to quantum mechanics itself. We are not going to do so, as it is outside the scope of this particular work, but we believe it is worth pointing it out.

In this section we saw how the quantum hypothesis requires us to abandon the classical description. Even ideal measurement, the best measurement we can possibly conceive, must be perturbative: the measurement of one quantity will change another. Such change makes the value after the measurement independent of the value before the measurement: future evolution cannot depend on the previous value, which also mean that it can no longer be measured. This independence of the initial value means that the state has a symmetry in the quantity: we can ideally change it yet future evolution is the same. A symmetry corresponds to a uniform distribution for the description of such quantity: if we measured it we would need to have an equal chance to get any value. Realizing how these elements are so tightly connected is the key, in our opinion, to understanding quantum mechanics: one cannot have one without the others, they are all different sides of the same coin. This way of presenting the subject does a better job, in our opinion, of bringing these concepts closer in the mind of the reader than what is found in standard textbooks.



\section{A dictionary}
Even if our aim is not to derive quantum mechanics we want to show that the concepts that we developed above find their place very naturally in the mathematical framework of quantum mechanics. We do this so as to not leave ambiguities, but also to show how disparate characteristics of the theory can find significance using the few concepts we have outlined. We will do this briefly in this section touching only the important points, as the full treatment of each particular item would essentially result in writing a textbook, which is not the purpose of this work either.

The notion of the ideal measurement translates, as the reader might expect, to the projection postulate: a measurement changes the state of the system to one of the eigenstates of the hermitian operator corresponding to the measurement, with the probability of doing so given by the projection of one onto the other. The eigenvalue is the outcome of the measurement and it is a real value, meaning that the result of the ideal measurement is exact. The projection is done \emph{at a particular time t}, meaning the ideal measurement is instantaneous. The state is changed during a measurement, so it is perturbative, but if the measurement is performed by more than one observer the final state will be the same, as further projections would yield the same result. It also does not depend on what process is used to measure. It satisfies all the properties we discussed.

But it has to be understood in terms of an \emph{ideal} measurement: a simplification, a limit, something that, in practice, we cannot do. What it is important is not really whether we can realize it in practice, but that \emph{each state can be seen as the outcome of this idealized process}. In mathematical terms, this means that for any state there exists a set of Hermitian operators that have that state as an eigenstate. This can be easily demonstrated to be true: let $\ket{\Psi_a}$ and $\ket{\Psi_b}$ be two normalized vectors in a Hilbert space $\mathcal{H}$, and let the first be an eigenket of Hermitian operator $A$. Given that they are normalized, we can always define a unitary operator $U$ such that $\ket{\Psi_b}=U\ket{\Psi_a}$: we do so by rotating on the plane defined by the two vectors and leaving the other directions unchanged. If we now consider the operator $B=UAU^{\dagger}$, it is easy to show that $\ket{\Psi_b}$ is an eigenket of $B$ and that $B$ is Hermitian.

In this light, \emph{the concept of state in both classical mechanics and quantum mechanics is the same}: the state is the outcome of a set of idealized measurements that must be specified to reproduce a system. The only thing that changes is that ideal measurements are perturbative in quantum mechanics. This creates the strongest possible connection between classical and quantum mechanics: it says how they are different and how they are the same.

The idea that one quantity is changed by measuring another is represented by conjugate quantities, such as position/momentum and spin/angle. The symmetry after a measurement corresponds to the unitary transformation generated by the Hermitian operator. An eigenstate of Hermitian operator $A$ will also be an eigenstate of the transformation $1 + A d\beta / i \hbar$, which means it is invariant under that transformation, and the distribution for an observable $B(\beta)$ that depends on parameter $\beta$ will be uniform, flat: \emph{uniform distributions and symmetries are the same thing}. With this in mind, it makes much more sense to say that \emph{final states of a measurement are eigenstates of the generated transformation}, since this actually has physical meaning, which is once again: for every quantity measured, or well defined, we have a symmetry. This also works in reverse: a state with a symmetry will have an associated observable well defined. This provides physical intuition for the Noether theorem: if a Lagrangian is invariant for a given transformation, the equations of motion preserve that symmetry and the associated observable.

Other important concepts can be understood in terms of just the ideas presented. Briefly: the commutator describes the change of one observable under a unitary transformation generated by a second, with the Heisenberg equation of motion being a specific example; a state with defined momentum is symmetric under translation, the evolution does not depend on position, hence non-locality; if the measured quantity spans two systems, such as the sum of their momentum, the symmetry also spans both systems, such as the sum of their position: we cannot describe the systems independently, the systems are entangled. This should give the reader a glimpse of how deeply these concepts are rooted in the theory.

As we said, we cannot go through an entire curriculum in this article. But we believe that we have at least shown how not only the concepts we used in the previous sections fit perfectly within the mathematical framework, but many more elements acquire a very straightforward significance if related to such ideas. This ability to explain the same number of concepts with fewer ideas can lead to a more elegant way to present the theory.

\section{Limitations of quantum mechanics}

We want to stress that the assumptions underneath ideal measurements in quantum mechanics are, well, ideal: at some point they will break down and need to be abandoned. To that end, we show in this section how the equivalence principle directly requires such a departure, which implies a limit of applicability for quantum mechanics. We do not intend this treatment to be exhaustive and it will be only in qualitative terms: our aim is just to show that problems exist and not to give a solution.

In our previous description of the ideal position measurement through the electromagnetic field we said that it is inter-independent because the laws that describe that field are linear. This is true, but we have not considered that each photon, having a finite energy, is going to interact with others through the gravitational force. This means that, if we introduce the equivalence principle, measurements affect each other and it will matter how many observers perform the same measurement at the same time: measurements are inter-dependent. If we want to describe the evolution of a system, we not only have to say what is measured at each time, but also, for example, what is the energy used as this will somehow affect the result of other measurements and future evolution in general: that information will have to be part of the state. It is also evident that the projection postulate of quantum mechanics is not at all suitable to describe an ideal measurement in which we consider the effect of gravity: it describes inter-independent measurements and is agnostic regarding the energy involved in the measurement.

Given that the gravitational force is orders of magnitudes smaller than the electromagnetic force, we do have cases in which the energy is such that the gravitational pull can be neglected (inter-independence), while the effect on the system being measured cannot (perturbation). This is the region where quantum mechanics will be valid. If we require higher energies, then we reach its limit. When is this the case?

From quantum mechanics itself we know that to probe shorter time scales we need to use higher energies: if the time scale required is "fast enough" the energy will be "large". Since instantaneous measurements would ideally require an infinite amount of energy, and infinite qualifies as large, we most likely need to abandon this assumption as well: we will need to specify what the duration of our idealized measurement is, which will tell us how much energy is needed. But we also need to note that, due to special relativity, the duration in time is connected to the precision in space. So our exact measurement assumption must also be abandoned.

The progression should be now evident: in classical mechanics we consider quantities as outcomes of completely ideal measurements; in quantum mechanics we start to consider the perturbation caused by measurements; here we see that to go further we also need to consider the strength, which also affects the precision in time and space. In other words, we are making our ideal measurement less and less ideal.\footnote{Note that the invariance of the speed of light, the quantum hypothesis and the equivalence principle all say something about ideal measurements: that is why their introduction always requires a new way of describing physical phenomenona.}

It is no surprise that quantum mechanics and general relativity are difficult to integrate. What is interesting is that the above argument is not \emph{technical} (as in whether the equations are linear or not) but it is \emph{conceptual} and pretty straightforward. Developing these ideas further is, naturally, not in the scope of this work, which is just to show how the quantum hypothesis requires us to abandon classical mechanics. What we did want to show is that the assumptions underneath quantum mechanics are, in a way, already known to be "wrong". As there is no "right" theory in physics, only ones that are accurate within the constraints of their validity, this section is meant to underline such constraints for quantum mechanics and to show how thinking in terms of idealized measurement allows one to do that very concisely.

\section{Discussion}

We feel that this work could end here. However, given that a lot has been written on quantum mechanics and many issues are still deemed as controversial, it may be important to better qualify our work so that it is at least clear what we do not expect it to be. Before one forms a definite opinion, we will discuss common reactions to these ideas and typical comparisons that have been made.

Some people, typically ones that only give only a superficial reading, argue that this work does not provide anything new: there is no new math developed, no new physical concepts, no new prediction. This shows a misunderstanding on our aim, which is to better understand a theory that is already now established for almost a century: it is unlikely that some important piece is missing. It is more likely, instead, that we do have all the pieces but we have not assembled them in a way that allows the big picture to emerge. Introducing new math or radical new physical concepts would be, in our opinion, either admitting our own failure or implying the failure of the people that came before us, and we are not holding our breath for the latter. \emph{The overall sense of familiarity that one may experience should be seen as an actual sign of strength of the argument}, as it lacks the implication that everyone has to review what he knows. And despite the familiarity, the arguments as presented are not in any work of which we are aware.

The differences are more in shifting importance between different elements to provide a better, more cohesive narrative. Such changes are subtle and we cannot expect each and every reader, having read this work once, to jump back in disbelief at how much their understanding has increased. The main reason being that \emph{the reader already knows quantum mechanics}. But if someone, as we did, uses this set of arguments to give a half an hour introduction on the subject to someone who is scientifically literate, you will discover that he can follow the thought experiments, he does not find the reasoning hard, is often convinced of the necessity of the arguments and of the striking consequences (like non locality and entanglement). This has not been our experience when we first learned quantum mechanics. This work can provide a mental scaffolding to students, who can get the big picture more quickly, and who will have a placeholder in which to put the rest of the content. If someone has already built the structure, it is clear the scaffolding is of less use. Some readers will still be intrigued by the simplicity and the elegance of the arguments, which to us has huge value in itself, but not everybody is bound to because, in the end, this judgement is subjective.

Some readers will try to put this work in the context of quantum mechanics reconstructions, which we can briefly describe as a set of different works that aim to derive the theory from a different set of postulates than the standard more mathematical ones.\cite{grinbaum} While we do believe that we could derive quantum mechanics from Einstein's quantum hypothesis, we are happy to simply note Rovelli's work and leave it at that: we feel that why we have to leave classical mechanics is as important, or possibly more, than where we have to go. When in dark nights we wonder about the mysteries of quantum mechanics, knowing that it is the consequence of "no bit commitment" does not comfort us (which may be due to our minimal background in quantum information theory). We want to know why we are forced to abandon classical mechanics. We also note that the choice of Einstein's quantum hypothesis as the foundational idea is not arbitrary: it makes the arguments work at multiple levels. It works because it is a clear physical statement of the world around us that is firmly experimentally established and is (at least now) non-controversial. It works because many feel that historically it is what started quantum theory in the first place. It works because it has been so much in front of our noses that it was difficult to notice its importance. It works because it allows to introduce quantum in a similar fashion to special and general relativity: through ideas and thought experiments. It works because it is what Einstein got his Nobel prize for. It \emph{feels} right. Even if we could not derive all of quantum mechanics from it, we personally do not feel the arguments would diminish in value. The fact that we could derive the mathematical framework as a whole is just the cherry on top.

Other readers will try to put this work in the context of interpretations. We can see how providing a dictionary from math to physics might make one think that this work belongs in that category. Here there are even more important distinctions, and we believe it instructive to compare and contrast with a couple of questions interpretations try to answer.

The first question we address is whether the wave function represents a real physical object or not. There is no claim in our work of what the wave function represents: in fact, there is no mention of the wave function at all. The state as a whole, though, is fully determined by a set of observables and their outcomes. This is also true during the evolution, so at each moment in time there is a set of observables that are well defined, possibly a different set every time. These observables are as real as real can get: they are defined whether we measure or not, their values are the same for all observers and their knowledge is enough to determine the state. This is the same as classical mechanics. What is different is that there are other observables, the conjugate ones, which do not determine the future evolution of the system and therefore cannot be measured. These, one might say, are not real at all: they are not determined a priori. The problem here, we believe, is simply the focus on valued measurement, and we hope the reader forgives this obvious statement: \emph{the symmetry in the conjugate variable is as measurable as the value of the original one}. The symmetry as a whole is as real as the valued quantities: it is there regardless of whether we measure it or not and it is the same for all observers. We always have a set of defined quantities and a set of symmetries no matter what, and it is obvious that the two are mutually exclusive for the same quantity. Time evolution or the act of measurement simply changes which quantity and symmetry. We let the reader decide whether this is enough to satisfy his thirst for realism, but it is for us.

Second question: can we have hidden variables? Most definitely: we can simply pick any value from the quantities in which we have a symmetry. There is nothing that says we should or we should not do so. The caveat is that, no matter what value we pick, it will never come into play in determining the future evolution of the quantities that fully determine the quantum state. If it did, then we would have a process that we can use to measure better than the ideal measurement, which is forbidden. That is why these variables are, in the end, hidden.\footnote{A hidden variable may only determine the future evolution of another hidden variable.}

The third question is the so-called measurement problem, which we briefly sum up. To put it simply, we have two ways of evolving a system: the Schroedinger equation for physical processes, which evolves the state continuously and is deterministic (given an initial state we can determine exactly the final state), and the projection postulate for measurements, which evolves the state discontinuously and is non-deterministic. The problem is that each measurement should be a process, but how can measurements be non-deterministic if all processes are deterministic?

To address this problem let us construct this thought experiment. We start at time $t$ in the eigenket $\ket{p_0}$ of $P$, and we perform an ideal measurement of some other quantity $A$. At time $t+dt$ we will be in an eigenket $\ket{a_i}$ of $A$, and we will perform an ideal measurement of $P$ so that at time $t+2dt$ we will be back in an eigenket of $P$. At the end will have a distribution in $P$, and by studying this distribution depending on what observable $A$ is we can understand a couple of things. First: what is the chance we are going back to the same eigenket? The probability to go from $\ket{p_0}$ to $\ket{a_i}$ is the same as for going from $\ket{a_i}$ to $\ket{p_0}$, and for the probability of going back we need to take into account all possible paths:
\begin{align*}
Prob(p_0, a_i) &= \braket{p_0}{a_i} \braket{a_i}{p_0} \\
Prob_{total} &= \displaystyle\sum_{a_i} Prob^2(p_0, a_i)
\end{align*}
If the $A$ corresponds to $P$, it is clear that we return to $\ket{p_0}$ with complete certainty: we never leave it. Let us consider the case where $A$ is not $P$, but one of its eigenstates $\ket{\hat{a}}$ is infinitesimally close to $\ket{p_0}$. In that case, we will change state but we are almost certain of our final state: determinism implies continuity and vice versa. We also require this change to be reversible, that is if we switch the two states, we get the same change but opposite in sign. Mathematically the two assumptions are:
\begin{align*}
\braket{\hat{a}}{p_0} &= 1 - \epsilon \\
\braket{p_0}{\hat{a}} &= 1 + \epsilon
\end{align*}
Under these condition, the probability to return is $1-\epsilon^2$: we return to the same state with certainty as we can disregard the second order infinitesimal term. But these are the same conditions under which the Schroedinger equation works: determinism, continuity and reversibility. Assuming reversible determinism implies continuous and unitary evolution, and from these two conditions we can actually \emph{derive} the Schroedinger equation!\footnote{See for example Sakurai\cite{sakurai}, pp 68-72. The derivation starts from probability conservation and continuity requirements.} So, in this special case, there is no actual problem. The more the eigenkets of $A$ differ, though, the more the jump will be non-deterministic and the final distribution in $P$ will be spread. The "worst" case is when $A$ corresponds to $X$, and we will have a uniform distribution in $P$.

But here we also have a bigger problem: if the momentum changes, the energy changes. Where is our system getting the energy from (or losing it to)? The only answer possible is: from the probe of our idealized measurement. Now think about it: if our intermediate measurement is $X$, the ideal measurement has to change the momentum to values that are possibly very very large. This means that the smallest amount of energy required by our ideal measurement is potentially infinite. But in the previous section we argued that if we \emph{require} energies that are large (and infinity still counts as large) we cannot use quantum mechanics. Whether and how these jumps do occur we still cannot say, but it seems clear that they cannot be described within quantum mechanics itself. Therefore it actually makes a lot of sense we do not have an equation describing such process of measurement: it would be nonsensical. It would describe a process based on the assumption that such process does not exist. This does not make quantum mechanics consistent, it simply shows at a conceptual level why such an inconsistency occurs.

So, it is true that this work may say something about the problems that interpretations try to address, but it is also true that it does not really provide answers: if anything it tells us the reason why we do not have them. Depending on the inclination of the reader, though, this may be far more interesting than the answers themselves, which is definitely the case for us.

We leave up to the reader to decide for himself what this work is closer to; we simply suggest that he keep these ideas in consideration. No matter what one may conclude, our claim is simply that this is a better way to introduce quantum mechanics.

\section{Conclusion}
We have presented a heuristic that can be used to conceptually justify the departure from classical mechanics, which we believe useful in the context of an undergraduate or graduate course of quantum mechanics. This material is complementary to the more standard ways of introducing quantum mechanics, because while it says from where we need to depart from and justifies some of the more puzzling aspects of quantum mechanics, it does not say where we need to end up.

The advantage of introducing the subject in this manner is that it brings it closer to the way that special and general relativity are usually introduced, using thought experiments that stimulate intuition, which we chose on purpose to be as similar as possible to those used for those theories. The other side effect of this introduction is the realization of how much is implied in classical mechanics: it is beneficial to remind students that the use of exact quantities, among other things, is an idealization. Though quantum mechanics is the theory that uses probability, it is classical mechanics that is in fact less precise and we believe this way of presenting reinforces that.

We also hope that this work is of interest to those who, like us, spent a significant amount of time reading a great number of textbooks in their youth trying to find a satisfying conceptual justification of quantum mechanics. May this work provide them at least part of it.

\end{document}